\documentclass[10pt]{article}
\usepackage{graphicx,amsmath}
\begin{document}
\title{Inflation and Global Equivalence}
\author{Geoffrey Martin \\
    Department of Mathematics \\
    University of Toledo \\
    Toledo, OH 43606      \\
    gmartin@math.utoledo.edu}
\date{}
\maketitle
\begin{abstract}
\noindent
 This article investigates an extension of General Relativity based upon a class of lifted metrics on the cotangent bundle
of space-time.  The dynamics of the theory is determined by a fixed section of the cotangent bundle, representing the
momentum of a fluid flow, and Einstein's equations for the fluid applied to the induced space-time metric on the
submanifold of the cotangent bundle defined by the image of the section. This construction is formally analogous to the
extension of Galilean Relativity by Special Relativity, and is shown to reduce to General Relativity as the
gravitational constant approaches zero. By examining the consequences of the model for homogeneous cosmologies, it is
demonstrated that this construction globalizes the equivalence principle, in that, the perfect fluid model of Special
Relativity is sufficient to predict both the inflationary and the current era.
\end{abstract}

\section{Introduction}
In the standard formulation of General Relativity (GR), it is axiomatic that frames are non-dynamical objects
determined by the choice of observers.  This feature of GR is a consequence of the clock hypothesis axiom, which
postulates that the rate at which an ideal clock (one that is corrected for the mechanical effects of acceleration)
operates depends only on its 4-velocity.  However, the role that frames play in dynamical structure of GR is more
complicated than suggested by their place in the axiomatic framework. This is because matter, as it is modeled in the
theory, often distinguishes a frame which becomes a dynamical object.  Examples of such frames are the 4-velocity
streamlines in fluid models and stationary observers in static space-times. The dependence of the space-time metric on
the distinguished frame of the matter flow suggests an extension of GR  in which the frame dependence of the metric is
incorporated into the axiomatic structure of the theory.

To see the possibility of such an extension, a frame of observers must be viewed as a geometric object in its own
right. This is best accomplished if a frame, rather than being viewed as a choice of coordinates on space-time, is
represented in an abreviated form, as a closed 1-form field $\lambda$ on space-time $M$. The data that $\lambda$
carries concerning the associated field of observers is the nature of the observers' clocks, and the local
decomposition of $M$ into space-like submanifolds. Now a 1-form field can also be viewed as a section of the cotangent
bundle $T^*M$ of $M$. From this perspective the geometric object most easily associated with a frame is its image
$\lambda(M)$.

To arrive at the proposed extension of GR, the frame of observers $\lambda(M)$ is thought of as an analogue of an
observer's world-line in Galilean or special relativity.  The fundamental problem in special or Galilean relativity is
to fix a representation of time along a world-line. A similar question arises in extending GR which is to establish
method for defining a space-time metric on $\lambda(M)$. The Galilean approach advocates an extrinsic procedure in
which there is a fixed correspondence that takes $\lambda(M)$ to a model space where distances are measured.  In
Galilean relativity the fixed extrinsic structure is the universal time line $\mathbf{R}$.  In the present case, $M$
and the bundle map $\pi\colon T^*M\to M$ serves this purpose. On the other hand, relativistic approach postulates that
distance measurement is intrinsic to the submanifold $\lambda(M)$ itself, and may depend on the position of the
submanifold in the ambient space.

The Galilean approach can be seen to yield the standard structure of GR. It is encoded in the axiom which for emphasis
Einstein states as {\it continuum spatii et temporis est absolutum} \cite{Ein}.  In this article we shall investigate
whether the relativistic approach may be the more powerful.  Following Einstein's dictum, if the relativistic approach
is to be a viable alternative, it must be demonstrated to be consistent, in the appropriate limit reduce to prior
tested theories, and have interesting consequences.  Consistency is of course a very difficult question in physics. The
second criteria of reduction to existing theoretical structure at least suggests a level of consistency similar to
existing theories. In following we hope to demonstrate that an extension of GR using the relativistic approach to
measuring space-time distances satisfies the last two of these criteria.

To show that the relativistic approach does in fact have interesting consequences, we shall examine the implications of
the extended theory for homogeneous cosmological models. Here, because in the extended theory the metric is frame
dependent, it is most natural to apply the equivalence principle in the matter frame.  This leads to a third rather
than second order equation for the cosmic scale factor; the solutions of which appear to have interesting global
consequences.  It will be shown that if the universe is assumed to evolve towards a dust state, then it must have
originated from an inflationary state. In fact the model suggests that the alternation of standard and inflationary
phases may be a cyclic process. Analysis of the model shows that it has the same consequences as current homogeneous
inflation models but with the advantage that cosmic evolution from the inflationary phase to the present can be
described as a single process.  The mathematical structure presented here does not yet apply to the study of
fluctuation in initial cosmic matter density, and so it can not be expected to account adequately for phenomena such as
reheating and condensation.

\section{Extended General Relativity}
In this discussion a frame on a space-time $M$ will be represented by the pair $(\lambda, V)$, where $\lambda$ is
1-form field on $M$, and $V$ is vector field on $M$.  A frame provides a decomposition of the tangent space to $M $ at
$p$ that is given by  $TM_p = \mathrm{span}(V_p)\bigoplus \ker(\lambda_p)$. This decomposition is interpreted as the
infinitesimal decomposition of $TM_p$ into space and time directions by the observer associated with the frame and
located at $p$. Note that this decomposition is independent of the choice of scale for the vectors $V$ and $\lambda$.
The scale of $V$ and $\lambda$ has useful descriptive properties.   The scale $V$ is associated with the observer's
choice of time coordinates and the scale of $\lambda$ can be interpreted as a mass scale for particles.

Although the association of 1-forms with mass is a standard correspondence in GR that is usually established from
momentum consideration, we provide here an argument for this relation based on Newton's Second Law. Suppose that $W$ is
another vector field on $M$, then the invariant velocity $w$ of $W$ in the frame $(\lambda, V)$ is given by
$w=(1/\lambda(W))W$.  Note that $w$ is invariant quantity in that it is independent of the choice of scale of $W$, and
hence is a geometric objects associated with the flow lines of $W$.  The velocity $w$, however, is dependent on the
choice of scale of $\lambda$. Newton's Law in the given frame may be expressed as $m\dot w = F(w)$, where $F$ is the
force and $\dot w$ is the invariant derivative of $w$.  The invariant derivative is given in terms of the Levi-Civita
connection $\nabla$ by the expression

\[\dot w = {1 \over\lambda(W)}(\nabla_W {W\over \lambda(W)})^\bot.\]

\noindent Here $\strut^\bot$ is the projection onto the complement of $V$.

Now suppose that the frame $(\lambda, V)$ is replaced by $(k\lambda, V)$ where $k$ is a positive constant. Let $w'$ be
the invariant velocity of $W$ in the frame $(k\lambda, V)$.  The above definitions imply that $m\dot {w'} = m(1/k^2)\dot
w = (1/k^2)F(w) = (1/k)F(w')= $ or $km\dot{ w'} = F(w')$.  Hence in the frame $(k\lambda, V)$ the mass of $W$ is
$m'=km$. Hence $\lambda$ can be imputed to have the same dimension as $m$

The equivalence principle guarantees that relativistic dynamics  can be reduced to Newtonian dynamics infinitesimally
along any flow line of $V$. However, it is only under very strong conditions that this correspondence can be extended
locally. In the case where $\lambda = m\ell(V)$, $\ell(V)$ being the 1-form metrically dual to $V$, the Newtonian
constructions gives an equivalent description of metric geometry only when $V$ is a gradient Killing field \cite{Mar3}.
This assumption implies that $V$ is holonomically trivial or $R(\cdot,\cdot)V=0$.  Unfortunately, a self gravitating
frame of massive observers cannot be holonomically trivial, and this fact leads to a global incompleteness in the
description of the physics of such systems. Mathematically, this difficulty surfaces in the lack of a globally
holonomic frame in which to apply the equivalence principle. This mathematical fact is mirrored in physical principle
that gravity destroys states of local thermodynamic equilibrium by invoking the other fundamental forces and thus
altering the local equation of state.  This is particularly true in cosmology where the evolution of fluid streamlines
can only be determined by fitting equations of state appropriate to each of the local temperature regime found in the
evolving universe.  The fact that there is no universal law that describes cosmic evolution thus can be traced to the
fact that we lack a universal description of the forces in nature. However, the fact that such a simple model as the
perfect fluid model appears to describe the bulk of cosmological space-time, suggests there should exist an underlying
principle from which a globally significant cosmological model can be derived.

Here we shall investigate the possibility that such a principle may be derived from a generalization of the equivalence
principle that arises in an extension of GR in which the space-time metric is frame dependent. This extension of GR is
obtained by a procedure that is analogous to the procedure used to obtain special relativity (SR) from Galilean
relativity (GLR).  Recall that the Galilean theory of time can be formulated by supposing that the space-time manifold
$M$ possesses a fibration $t\colon M \to \mathbf{R}$. This fibration is assumed to possess a connection represented by a
vector field $V$ on $M$ which must necessarily have the property that if $\phi_s$ is the flow of $V$ then
$\phi_{s_1}(t^{-1}(s_2)) = t^{-1}(s_1+s_2)$.  Thus the pair $(dt,V)$ determines a synchronous field of observers on $M$
with the special property $dt(V)=1$.  Let $N$ be a model for the fiber of $t$. GLR supposes that $N$ and $\mathbf{R}$
are separately endowed with the complete Riemannian metric tensors $q$ and $g$, and that $g$ is solely responsible for
determining the time between two events $p, q \in M$. If $\delta_g$ is the topological metric on $\mathbf{R}$
determined by $g$, then according to GLR, the time between $p$ and $q$ is given by $\delta_g(t(p),t(q))$. In SR time
does not have this well defined character but depends on the observer with respect to which these events are
referenced. The metric that an observer uses is induced on the observer world line from a metric $\widetilde g$ on $M$
that is obtained by summing $g$ and $q$. This is done using the decomposition $TM_p = \mathrm{span}(V_p)\bigoplus
\ker(dt_p)$, the natural isomorphisms $t_*\colon \mathrm{span}(V)\to T\mathbf{R}$, and the identification $i_s\colon
t^{-1}(s)\to N$. The metric $\widetilde g$ is then given by the expression $\widetilde g = -t^*g\bigoplus (1/c)^2i^*q$
where $c$ is a universal constant. In the limit as $c\to \infty$, SR reduces to GLR.

This procedure can be carried out again with the base manifold $\mathbf{R}$ replaced by the space-time manifold $M$, and
the fibration $t\colon M \to \mathbf{R}$ replaced by a fibration $\pi\colon\overline{M}\to M$. The manifold
$\overline{M}$ that replaces $M$ can be discovered by noting that in SR the fiber $t^{-1}(s)$ parameterizes the
positions of all possible observers at time $s$.  Thus, by analogy it is natural to suppose that for an event $p\in M$
the fiber $\pi^{-1}(p)$ parameterizes the frames of all possible observers at $p$.  The above description of frames
indicates that the fiber should be $TM_p\times T^*M_p$. However, it will be assumed that there is a relation
$\ell\colon TM\to T^*M$, and that the frames under consideration all have the form $(\lambda,\ell^{-1}(\lambda))$, and
thus the fiber of $\pi$ at $p$ may be assumed to be $T^*M_p$, making $\overline{M}=T^*M$.

As in SR, a metric can be constructed on $T^*M$ using the Lorentzian metric on $M$ and the induced metric on the fibers
of $T^*M$.  To interpret this construction it is helpful to recall how relativity unifies space and time dimensions. In
the case of SR, tangent vectors to $N$ are assigned the dimension of length $\mathsf{L}$ while tangent vectors to
$\mathbf{R}$ are assigned the dimension of time $\mathsf{T}$.  Metric tensors are supposed to be dimensionless. If the
universal constant $c$ is supposed to have the dimension $\mathsf{L/T}$, then for vector fields  $V$ and $W$ on $M$,
$\widetilde g(V,W)$ has dimension $\mathsf{T}^2$. This suggests that the dimensional structure of $M$ should be
simplified by measuring the length of tangent vectors to $N$ in $\mathsf{T}$-units with 1 $\mathsf{L}$-unit equaling
$1/c$ $\mathsf{T}$-units. In such a system $c=1$ and all physical tangent vectors to $M$ have dimension $\mathsf{T}$.

 In the case of $T^*M$, the above discussion of Newton's Law suggests that the fibers
of $T^*M$ should be given the dimension mass $\mathsf{M}$. Thus in this case we see that there is the need of a
physical constant $1/G$ of dimension $\mathsf{M/T}$ relating the base and fiber dimensions. The constant $G$ is
easily seen to have the same dimension as the gravitational constant.

To construct a metric on $T^*M$ denote the space-time metric by $g$ and proceed analogously to construct a metric
$\widetilde g$ on $T^*M$. In analogy with GLR, the horizontal distribution $H$ of the Levi-Civita connection
plays the role of the  field $V$. It is a well known fact in connection theory that $TT^*M$ decomposes as
$TT^*M_p=H_p\bigoplus VT^*M_p$, where $VT^*M$ is the vertical distribution of the fibration $\pi\colon T^*M\to M$.
Because $T^*M$ is a vector bundle there is a natural isomorphism $i_p\colon V_p \to T^*M_{\pi(p)}$ defined for
each $p\in T^*M$.  The isomorphism $i_p$ together with the isomorphism ${\pi_*}_p\colon H_s \to TM_{\pi(p)}$ can
be used to define a class of metrics on $TT^*M$ with the property that $H$ and $VT^*M$ are orthogonal
distributions. Of particular interest is the metric $\widetilde g$ given by
\[
\widetilde g_p = -{\epsilon\over g(p,p)}(i_p)^*g\bigoplus {g(p,p)\over \epsilon}({\pi_*}_p)^*g.
\]

This expression is considerably more complicated than the corresponding expression in SR due to the presence of the
metric dependent conformal factors. There are several reasons why this particular form was chosen for $\widetilde g$.
First, the parallelism of this metric can be shown to be closely related to Fermi parallelism along curves on the
space-time \cite{Mar2}. Thus, the relativistic effects of acceleration are encoded in the geometric structure  defined
by $\widetilde g$ \cite{Mar1}. Second, since $\widetilde g(X,Y)$ should have a well defined dimension, the dimensions of
the vertical and horizontal summand of this quantity should be equal.  This implies that ${\epsilon/g(p,p)}$ has
dimension of the gravitational constant $\mathsf{T/M}$, and so $\epsilon$ has the dimension of Plank's constant
$\mathsf{MT}$. Thus the dimension of the conformal factors can be factored as $\mathsf{MT}/\mathsf{M}^2$.  This
factorization is similar to the factorization of the gravitational constant $G=\hbar/{\mathsf{M}_P}^2$ where
$\mathsf{M}_P$ is Planck's mass. This particular way of expressing the dimensional dependence of $\widetilde g$ will
prove very useful in cosmological applications.

To interpret the metric $\widetilde g$, consider a field of observers associated with a section $\lambda$ of
$T^*M$. Just as in SR where the world-line of an observer inherits a metric that measures time displacement, in
this case the image of space-time $\lambda(M)$ inherits a metric $G\widetilde g|_{\lambda(M)}$ which measures the
space-time displacement in that frame. Here a dimensional constant $G$ is introduced with dimension
$\mathsf{T}/\mathsf{M}$ so that when evaluated $G\widetilde g$ has dimension $\mathsf{T}^2$. Thus in extended
general relativity (EGR), an observer field $\lambda$ detects not the space-time metric $\pi^*g$  but rather the
space-time metric $G\widetilde g|_{\lambda (M)}$. It will be demonstrated in section 3 that if $G\to 0$ and
$\epsilon\to 0$ in such a way that $Gm^2/\epsilon \to 1$, then EGR reduces to GR.

\section{Global Equivalence in Friedmann Geometry}

This section shall examine some consequences of EGR in the standard cosmological model.  In this case, space-time
$M$ is modeled by the product $\mathbf{R}\times N$ where $N$ is now a simply connected 3-dimensional space-form of
constant sectional curvature $\kappa$.  If $q$ is the constant curvature metric on $N$ and if $t$ is the natural
affine coordinate on $\mathbf{R}$, then consider the class of Friedmann metric tensors
\[
g=-dt^2\bigoplus S(t)^2q,
\]
where $S(t)$ is assumed to be a smooth positive function.  Freidmann geometries possess a distinguished frame
determined by the unit vector field  $T$ in the distribution $T\mathbf{R}\times 0$. This frame is  described by
the pair $(T,s)$ where $s=m\ell(T)$ and $m$ is some intrinsic mass.  To study the geometry of the frame $(T,s)$
first observe that $\widetilde g|_{s(M)}$ is of Robinson-Walker type. To determine the form of $\widetilde g$ on
$s(M)$ use the induced map $s_*\colon TM\to Ts(M)$ to parameterize $Ts(M)$. It follows from the theory of linear
connections that for any tangent vector $u\in TM_p$,
\[
s_*u = \widetilde u_{s(p)}+{i_{s(p)}}^{-1}\nabla_us,
\]
where $\nabla$ is the dual Levi-Civita connection of $g$ and $\widetilde{\strut}$ denotes the horizontal lift to
$TT^*M$ determined by $\nabla$. From the identities
\begin{eqnarray*}
\nabla_Ts&=&0 \\
\nabla_Xs&=&m\dot{\log S(t)}\ell(X),
\end{eqnarray*}
where $X$ is a vector field in the distribution $0\times TN$, and the above expressions for $s_*u$ and
$\widetilde g$, it is easy to see that
\begin{eqnarray*}
\widetilde g(s_*T,s_*T)&=&{m^2\over \epsilon}\\
\widetilde g(s_*X,s_*Y)&=&-{m^2\over \epsilon}(1-{\epsilon^2\over m^2}(\dot{\log S(t)})^2)S(t)^2q(X,Y).
\end{eqnarray*}
These expressions imply that since $\pi\colon s(M)\to \mathbf{R}\times N$ is a diffeomorphism
\[
\widetilde g|_{s(M)}={m^2\over\epsilon}(\pi^*dt^2\bigoplus -(1-{\epsilon^2\over m^2}(\dot{\log
S(t)})^2)S(t)^2\pi^*q).
\]

Since the intrinsic geometry of the frame determined by $s$ is a Friedmann geometry, it provides a  model of a self
gravitating perfect fluid.  To complete the model one needs in addition to Einstein's equations an equation of state
for the fluid. At this point it is customary in GR to appeal to the equivalence principle and to  SR for an equation of
state.  In EGR, since the metric is frame dependent, there is the question of choosing the frame in which to apply the
equivalence principle.  The hypothesis that shall be advanced is that the equivalence principle should be applied in
the frame of the matter flow, and that in this frame it yields an equation of state with global validity. In other
words, the entropy generating effects of gravity can be eliminated by "viewing" space-time from the matter frame. This
idea can be thought of as an extension of Einstein's original version of the equivalence principle in which
gravitational forces were eliminated by accelerated frames. The original form of the equivalence principle was too
strong as it only applies infinitesimally or to pseudo-gravitational forces as note above. However, because the metric
in EGR is frame dependent the original form of the equivalence is no longer limited to flat geometries and appears to
have interesting implications.

To derive the consequences of this assumption recall that for a simple gas, kinematic analysis implies that the
energy $e$, pressure $p$ and density $\rho$ satisfy the equations of state $e-\rho=\alpha p$ and that the energy
momentum tensor has the form $T= (e+p)V\otimes V + pg$. When coupled with Einstein's equations, the equation of
state yields a system of equations that determine $S(t)$. For simplicity let
$S'(t)^2=(1-(\epsilon^2/m^2)(\dot{\log S(t)})^2)S(t)^2$.  From Einstein's equations in the frame determined by
$s$, we have
\begin{eqnarray*}
e&=&{3\over G}((\dot{\log S'(t)})^2+{\kappa\over S'(t)^2}) \\
p&=&-{1\over G}(2\ddot{\log S'(t)}+3(\dot{\log S'(t)})^2+{\kappa\over S'(t)^2}).
\end{eqnarray*}
Further the matter conservation law implies that there is a constant $g$ such that $\rho = g/S'(t)^3$.

To obtain the equations for $S(t)$ implied by these relations and the equation of state, first assume that, as
suggested in the last section, $G=\epsilon/m^2$.  It will also be useful to transform to the universal time scale
by setting $u(t)=S(kt)$ where $k= \epsilon/m$.  With these assumptions a calculation shows that $u(t)$ satisfies
\begin{eqnarray*}
\lefteqn{\dot{\log u}(1-\dot{\log u}^2)\dddot{log u} =}\\
            & &\ddot{\log u}(1-\dot{\log u}^2)(1-{3+4\alpha\over \alpha}\dot{\log u}^2)- \ddot{\log
            u}^2(1-{3+\alpha\over 2\alpha}\dot{\log u}^2) + \\
            & &{3(1+\alpha)\over 2\alpha}\dot{\log u}^2(1-\dot{\log
            u}^2)^2 + {3+\alpha\over 2\alpha}{\kappa k^2\over u^2}(1-\dot{\log u}^2) + {1\over 2\alpha}{k^2Gg\over
            u^3}(1-\dot{\log u}^2)^{1\over 2}.
\end{eqnarray*}
It is physically reasonable to assume that the constants $(3+\alpha/ 2\alpha)\kappa k^2$ and $(1/2\alpha)k^2Gg$
are both extremely small, and so the last two terms in the above expression can be neglected.  If these terms are
dropped then the above equation for $u$ can be rewritten as a planar system in the variables $x(t)=\dot{\log
u(t)}$ and $y(t)=\ddot{\log u(t)}$ which after rescaling time has the form $\dot{x}=F_x(x,y)$ and $\dot y =
F_y(x,y)$ where
\begin{eqnarray*}
F_x &=&x(1-x^2)y \\
F_y&=& y(1-x^2)(1-{3+4\alpha\over \alpha}x^2)-y^2(1-{3+\alpha\over 2\alpha}x^2)+{3(1+\alpha)\over
2\alpha}x^2(1-x^2)^2.
\end{eqnarray*}

This system, although it can not be explicitly solved, can be analyzed using the qualitative theory of planar systems.
To interpret this analysis it should be kept in mind that the $xy$-phase plane is related through Einstein's equations
to the Hubble constant-pressure plane of the cosmological model.  First note that the above system possesses four
stationary points $(0,0)$, $(-1,0)$, $(1,0)$ and $(0,1)$.  The stationary point $(0,1)$ is easily seen to be of
hyperbolic type. However the analysis of this equation is complicated by the fact that $(0,0)$ is semi-degenerate and
$(-1,0)$ and $(0,-1)$ are totally degenerate.  The stationary point (0,0) will be shown to be topologically of
hyperbolic type and the stationary point at $(0,1)$ will be shown to possess elliptic and parabolic sectors. Using the
symmetry of the equation it may be assumed that $x\geq 0$. This is equivalent to assuming that the Hubble constant is
positive.

To understand the global behavior of solutions to this system first consider the structure of solutions near
$(0,0)$. Since $F_x(0,y)=0$, it follows that $x=0$ defines a solution curve that adheres to $(0,0)$. To find the
other adherent solution invoke polar coordinates by the standard relations $\dot r=F_r(r,\theta)$ and $r\dot\theta
= F_\theta(r,\theta)$ where $F_r(x,y)=(xF_x(x,y)+yF_y(x,y))/r$ and $F_\theta(x,y)=(xF_y(x,y)-yF_x(x,y))/r$. The
other adherent solution can be located by considering a sector bounded by $y=0$ and a solution of $F_r(x,y)=0$.
Note that the equation $F_r(x,y)=0$ is quadratic in $y$ and may be solved to give radial isoclines
\[
y_\pm={(1-x^2)(1-{3+4\alpha\over \alpha}x^2)\pm\sqrt{(1-x^2)x^4({3+7\alpha\over \alpha}-12{1+\alpha\over
\alpha}x^2)+(1-x^2)^2(1-x^2)^2}\over 2(1-{3+\alpha\over 2\alpha}x^2)}.
\]
We may distinguish radial isoclines  by the property that $y_+(0)=1$ and $y_-(0)=0$.  Let $I_0$ be the graph of $y_-$
for $x\geq 0$. Consider a small sector $W$ bounded by $y=0$ and $I_0$. The vector field $F$ exits both radial
boundaries of $W$. First, because $F_x(x,0)=0$ and $F_y(x,0)\geq 0$, $F$ exits $y=0$. Second, a Taylor expansion for
$F_\theta(x,y)$ developed along $I_0$ implies that $F_\theta(x,y_-(x))=-x^3+o(x^4)$, and consequently, if the sector
$W$ is chosen sufficiently small, $F_\theta\leq 0$ on the segment of $I_0$ bounding $W$. As a result, there must either
exist a unique solution or an interval of solutions that tend to $(0,0)$ as $t\to\infty$ with initial conditions on the
arc bounding $W$.

To see that there is a unique solution, examine the first order ode $r(d\theta/dr)=\Psi(x,y)\equiv
F_{\theta}(x,y)/F_r(x,y)$. Introduce the angular operator $\Theta=(x\partial/\partial y-y\partial/\partial x)/r$.
It follows from a generalization of a lemma of Lonn that there is a unique solution adhering to $(0,0)$, if
$\Theta\Psi<0$ in the interior of $W$ \cite{Nem}. To verify this inequality, note that
$\Theta\Psi(x,y)=Q(x,y)/F_r(x,y)^2$, where $Q(x,y)$ is a polynomial of order 14 in $x$ and $y$ with properties
that $Q(x,y)=-4\alpha y^2(x^2+y^2)+o(r^4)$ and $Q(x,0)=-3(5\alpha+3)(\alpha+3)x^6+o(x^7)$. These facts imply that
$Q(x,y)$ is negative definite in a neighborhood of the origin, which implies that if $W$ is chosen small enough,
then $\Theta\Psi<0$ on the interior of $W$.  Hence, there is a unique solution that adheres to $(0,0)$ in $W$,
and determines a stable manifold $J_0$ for the stationary point $(0,0)$.

To understand the global backward time evolution of $J_0$, first consider the isoclines $F_y(x,y)=0$. This
equation is again quadratic in $y$ and has the solutions
\[
\tilde y_\pm=(1-x^2)\left ({(1-{3+4\alpha\over\alpha}x^2)\pm\sqrt{12{\alpha+1\over\alpha}x^4+(1-x^2)^2}\over
2(1-{3+\alpha\over 2\alpha}x^2)}\right ).
\]
These isoclines again can be distinguished by $\tilde y_-(0)=0$ and $\tilde y_+(0)=1$. The most crucial feature of the
graphs of $\tilde y_-$ and $\tilde y_+$ is the position and sign of the vertical asymptotes. It can be seen that only
$\tilde y_-$ possesses an asymptote  at $x_0=\sqrt{2\alpha/(3+\alpha)}$ while $\tilde y_+$ is finite at $x_0$ and has
the value $\tilde y_+(x_0)=(3(3-\alpha)(1+\alpha))/(4(3+\alpha)(7\alpha+3))$. The sign of the asymptote depends on
whether $x_0>1$ or $x_0<1$. It is easily seen that $\lim_{x\to x_0^-}y_-(x)=-\infty$ if $x_0<1$ and $\lim_{x\to
x_0^-}y_-(x)=\infty$ if $x_0>1$. Also note that both isoclines have the property that $y_\pm(\pm1)=0$.

\vspace{.5cm} \makebox[4.5in]{\includegraphics[width=350pt,bb=32 252 547 576,clip=true]{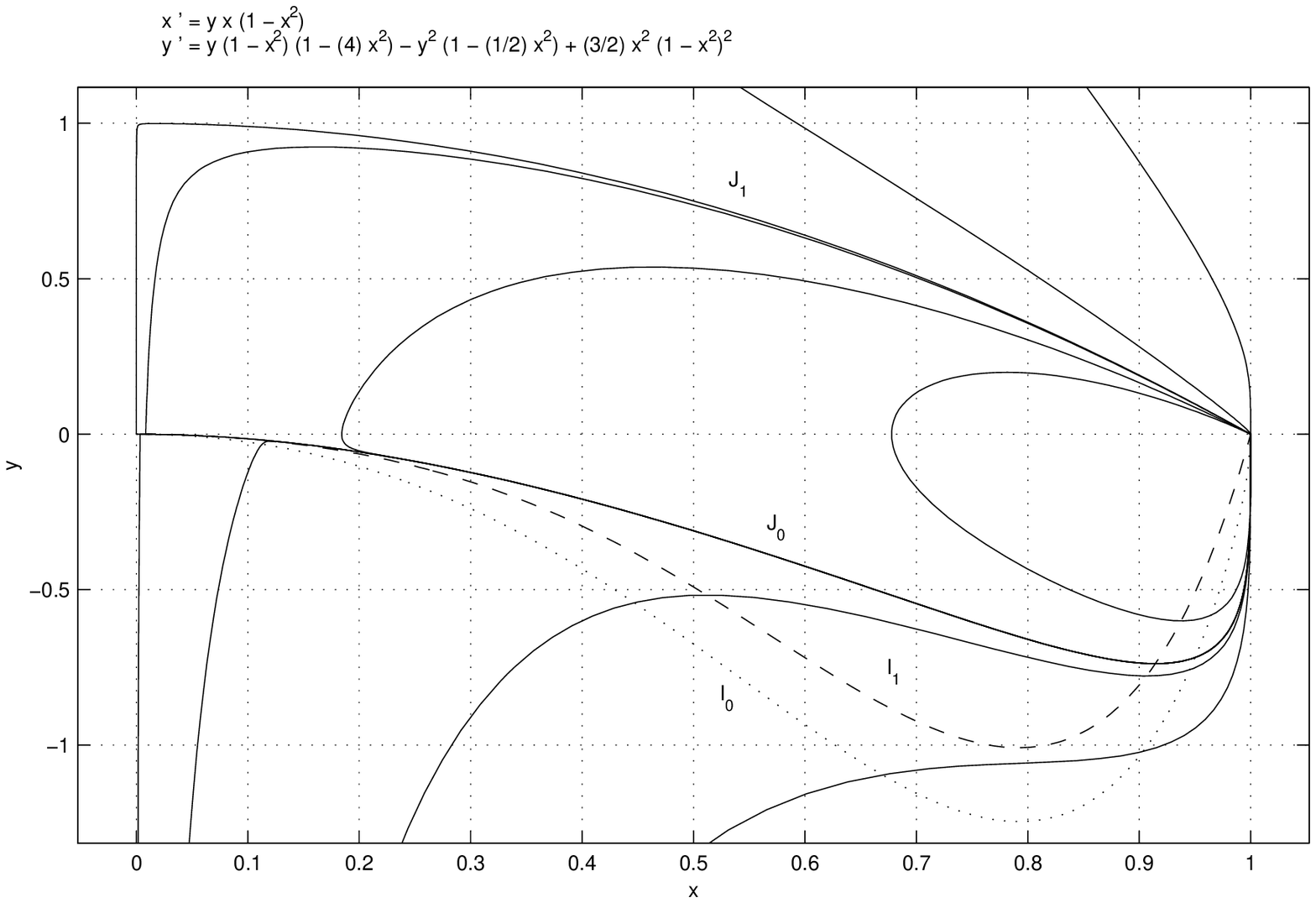}}
\begin{center}
\parbox{4.6in}{\small Figure 1. Flow lines for the equation $\dot x= yx(1-x^2)$, $\dot y =
y(1-x^2)(1-4x^2)-y^2(1-(1/2)x^2)+(3/2)x^2(1-x^2)^2$ showing stable manifolds $J_0$ and $J_1$ and radial and cartesian
isoclines $I_0$ and $I_1$.}
\end{center}
\vspace{.4cm}

To understand the relation between the stable manifold $J_0$ and the isocline $I_1$ determined by the graph of $\tilde
y_-$, we develop a Taylor expansion for $J_0$ at $(0,0)$. We may assume that $J_0$ can be parameterized as the graph of
a smooth function $f(x)$; where $f$ must satisfy $F_y(x,f(x))=f'(x)F_x(x,f(x))$. To obtain the first two terms in the
Taylor expansion, assume that $f(x)=Cx^2+Dx^4+o(x^5)$ and substitute into the above expression. It can be seen that
$f(x)=-(3(1+\alpha))/(2\alpha)x^2+(3(1+\alpha))^2/(2\alpha)^2x^4+o(x^5)$ while a Taylor expansion for $\tilde y_-$ at
$(0,0)$ has the form $\tilde y_-(x)= -(3(1+\alpha))/(2\alpha)x^2-(3(1+\alpha))^2/(2\alpha)^2x^4+o(x^5)$.  Thus the
stable manifold $J_0$ lies above isocline $I_1$.


Consequently, the global behavior of $J_0$ is determined by the position of the asymptote $x_0$. If $\alpha<3$ or
$x_0<1$, $J_0$ lies in an unbounded region defined by $y=0$, $x=1$ and $I_0$ in which case $F_y>0$ and $F_x>0$.
Since $x=1$ is a solution curve $J_0$ cannot intersect it , and thus is asymptotic to it as $t\to -\infty$. The
case where $\alpha>3$ or $x_0>1$ is quite different.  Here $J_0$ lies in a bounded region $U$ defined by $y=0$ and
$I_1$ (see Fig. 1). In this case divide $I_1$ into two segments $I_1'$ and $I_1''$ at the point $(x_1,y_1)$ where
$I_1$ attains its minimum $y$-value. Assume $(0,0)\in I_1'$.  It is easy to see that the vector field $F$ exits
$U$ along $y=0$ and $I_1'$ and enters along $I_1''$. Consequently, in backward time $J_0$ either adheres to the
stationary point $(1,0)$ or leaves $U$ along $I_1''$. In latter case since $F_y(x,y_1)<0$ for $x_1<x<1$, and
since $x=1$ is a solution, it follows that, if $J_0$ enters the region bounded by $I_1''$, $x=1$ and $y=y_1$, it
must adhere to $(1,0)$.  Thus in the case where $\alpha>3$ the manifold $J_0$ is an unstable manifold for $(1,0)$
and a stable manifold for $(0,0)$.

The unstable manifold $J_1$ leaving the stationary point $(0,1)$ can also be seen to adhere to (1,0). This follows
from the fact that the boundary of the unit square $[0,1]\times [0,1]$ either consists of solutions curves, $x=0$
and $x=1$, or segments where the vector field is entering. Since $J_1$ enters the unit square, it must adhere to
a singular point on the boundary. However, it cannot adhere to $(0,0)$, since the only inward pointing
characteristic direction with angle between $0$ and $\pi/2$ is uniquely associated with $J_0$. Thus $J_1$ must
adhere to $(1,0)$.

The final step in understanding the global behavior of the flow of $F$ is to analyze the stationary point
$(1,0)$. Here we shall restrict our attention to the case where $\alpha\to\infty$; that is to zero pressure
cosmologies.  In this limit we find that
\begin{eqnarray*}
F_x &=&x(1-x^2)y \\
F_y&=& y(1-x^2)(1-4x^2)-y^2(1-{1\over 2}x^2)+{3\over 2}x^2(1-x^2)^2.
\end{eqnarray*}

To study the stationary point at $(1,0)$ introduce the coordinate change $z=1-x$ and transform to polar
coordinates letting $z=r\cos(\theta)$ and $y=r\sin(\theta)$. After the standard rescaling of time, one obtains the
equation for the blow-up of the vector $F$ at $(1,0)$; namely $\dot r = R(r,\theta)$ and $\dot \theta =
S(r,\theta)$ where $R(r,\theta)=rR_0(\theta)+\eta(r,\theta)$ and $S(r,\theta)=S_0(\theta)+\xi(r,\theta)$ with
$\lim_{r\to 0}\xi(r,\theta)=0$ and $\lim_{r\to 0}\eta(r,\theta)/r=0$. The characteristic directions are
determined by the equation $S_0(\theta)=0$.  The solutions to this equation are easily obtained since
\begin{eqnarray*}
S_0(\theta)&=&{15\over 4}\cos(\theta)(1+\cos(2\theta+\varphi)) \\
R_0(\theta)&=&{15\over 4}\sin(\theta)({7\over 15} + \cos(2\theta+\varphi))
\end{eqnarray*}
where $\varphi = \arctan(4/3)$. The equation $S_0(\theta)=0$ is easily seen to have solutions $\pi/2$, $-\pi/2$,
$(\pi-\varphi)/2$ and $(\pi-\varphi)/2+\pi$. Also a short calculation gives $(\pi-\varphi)/2=\arctan(2)$.

To determine whether there are solutions to the ode adhering to $(1,0)$ along these characteristic directions,
first note that at any of these angles $R(\theta)\neq 0$. Consequently, the existence of such solutions is
determined by the Taylor expansion of $S_0$ at the given characteristic angle. In general, if the first
non-vanishing coefficient is of odd degree then there exist solutions to the ode adhering along that
characteristic direction. However, if the first non-vanishing coefficient is even then there exist either no
solutions or an interval of solutions adhering along that characteristic direction. Examination of the above
expression for $S_0$ shows that the Taylor expansions of $S_0$ at $\pi/2$ and $-\pi/2$ are of odd degree, and at
$\arctan(2)$ and $\arctan(2)+\pi$ they are of even degree. In both cases there are solutions adhering to $(1,0)$
along these characteristic directions \cite{Har}.

Any solution with initial conditions $(x,0)$ for $0\leq x\leq 1$ must adhere in forward time to $(1,0)$ along the
characteristic direction $\arctan(2)$ or $\pi/2$, and in backward time along the characteristic direction
$-\pi/2$. Since solutions adhere to $\arctan(2)$ exist, there must exist a closed interval $[x_0,1]$ such that if
$x\in [x_0,1]$ then solutions with initial conditions $(x,0)$ adhere to $(1,0)$ along the $\arctan(2)$ direction.
Numerical results suggest that in fact $x_1=0$, and that the last solution adhering to $(1,0)$ at the angle
$\arctan(2)$ is $J_1$. Hence it is conjectured that the solution curves $x=0$ for $0\leq y\leq 1$, $J_0$ and
$J_1$ form a separatrix for an elliptic sector of the stationary point $(1,0)$.

To develop a physical interpretation for these results, first note that the rescaling $u(t)=S(kt)$ and Einstein's
equations imply that points in $xy$-plane are related to the energy and pressure of cosmological states by the
relations
\begin{eqnarray*}
e&=&{m^4\over\epsilon^2}3x^2\\
p&=&-{m^4\over\epsilon^2}(2y+3x^2).
\end{eqnarray*}

The most important integral curve of the vector field $F$ for the interpretation of these results is $J_0$. It not only
approximates the standard cosmological model for large time, but also, in the case where $\alpha \geq 3$, it originates
from an inflationary state. To see this, rescale vector field $F$ so that its parameter correctly reflects the time
evolution of the original problem; that is $\dot x = y$. This rescaling can be accomplished by representing the
solution curve as the graph of $f(x)$, and then solving for $x(t)$ by integrating $\dot x = f(x)$. In the case of $J_0$
at $(0,0)$, from above, we know that $f$ has the form $f(x)=-((3+3\alpha)/2\alpha) x^2+g(x)$ where $\lim_{x\to
0}g(x)/x^3=0$. Consequently, integration gives that $x(t)=(2\alpha/(3+3\alpha))(1/t)+b(t)$ where
$\lim_{t\to\infty}b(t)=0$.  But, this is just the asymptotic behavior of solutions to the standard model with sectional
curvature $\kappa = 0$ and density $\rho \ll 1$. Thus $J_0$ or solutions close to $J_0$ are candidates for equations of
state that give reasonable cosmologies. Following $J_0$ in backward time, it either originates from a state of infinite
pressure in the case where $\alpha<3$, or a state where $p=-e$ in the case where $\alpha \geq 3$.  The latter equation
of state is the equation for inflationary cosmology. The fact that for $\alpha > 3$, cosmological models describe cold
matter weakly interacting with radiation indicates that, in this scenario, the inflationary phase is predicted by the
current condition of the universe. Note that the energy density at the singularity is $3m^4/\epsilon^2$. If $m$ is
taken to be the Planck mass this is consistent with the chaotic inflation hypothesis.

To gain a better insight into the significance of this model for inflationary cosmology, requires a more detailed
examination of the solutions near the stationary point $(1,0)$.  Using a similar analysis as was applied at
$(0,0)$, it can be shown that solution curves that adhere to $(1,0)$ are the graphs of either $f(z)=
-z^{1/4}C(z)$ where $C$ is analytic and $C(0)>0$ or $g(z)=2z+h(z)$ where $\lim_{z\to 0}h(z)/z=0$. These solutions
correspond to the $-\pi/2$ and $\arctan(2)$ characteristic directions respectively. To find $x(t)$ solve $\dot z
= -f(z)$ and set $x=1-z$. In first case with the initial condition $(t_0,z_0)$, integration gives for $0<z<z_0$,
$z^{3/4}\widetilde C(z)-z_0^{3/4}\widetilde C(z_0)=t-t_0$ where $\widetilde C(0)= 4/(3C(0))$. Let
$u^{4/3}D(u^{4/3})$ be the inverse of $z^{3/4}\widetilde C(z)$ with $D(0)=1/\widetilde C(0)^{4/3}$, then the
solution $x(t)$ has the form
\[
x(t)= 1-((t-t_0) + z_0^{3\over 4}\widetilde C(z_0))^{4\over 3}D(((t-t_0) + z_0^{3\over 4}\widetilde
C(z_0))^{4/3}).
\]
One significant observation that follows from this expression is that, if $J_0$ is parameterized by the natural
time parameter, then the stationary point is reached along $J_0$ in finite time. In fact, since $z_0^{3\over
4}\widetilde C(z_0))>0$ the singularity is reached at time $t_1<t_0$, $t_1=t_0-z_0^{3/ 4}\widetilde C(z_0)$.
Knowing $x(t)$ we may also calculate the scale factor $u(t)=\dot x(t)/x(t)$. Integration gives
\[
u(t) = u(t_1)\exp((t-t_1)-((t-t_0)+z_0^{3\over 4}\widetilde C(z_0))^{7\over 3}\widetilde D(((t-t_0)+z_0^{3\over
4}\widetilde C(z_0))^{4/3}).
\]
The analysis of this expression shows that, although for small $t-t_1$, $u(t)$ grows exponentially, exponential
growth does not persist long enough to account for inflation.  To find true inflationary growth in this model we
examine the behavior of solutions adhering to $(1,0)$ along the characteristic direction $\arctan(2)$.

For solutions adhering along the $\arctan(2)$ direction, the integral of $\dot z = -g(z)$ can be expressed in
terms of the initial data $(t_0,z_0)$ and a bounded function $b(z)$, and has the form
$zb(z)=z_0b(z_0)e^{-2(t-t_0)}$ for $0<z<z_0$. If the inverse of $zb(z)$ is $vd(v)$, where $d$ is also bounded,
then $x(t)$ has the form
\[
x(t)=1-z_0b(z_0)e^{-2(t-t_0)}d(z_0b(z_0)e^{-2(t-t_0)}).
\]
Hence the scale factor $u(t)$ has the form
\[
u(t)=\exp((t-t_0)+z_0b(z_0)e^{-2(t-t_0)}\tilde d(z_0b(z_0)e^{-2(t-t_0)})).
\]
From this expression we see that as $t\to \infty$ the scale factor undergoes exponential growth.  This fact
suggests that the solutions of the elliptic sector may be given physical interpretation if it is supposed that the
inflationary phase of the universe's development was the final state of a previous cycle.

As final remarks, observe that both the flatness of space and the adherence to the critical density can be deduced
from this model, but in a manner quite different from the way in which these conclusions arise in standard
inflationary models. First, the fact that the density of matter must be extremely close to critical density is a
consequence of the age of the universe relative to universal time scale $k$.  Since, if cosmic evolution did not
follow a path extremely close to $J_0$, it would have suffered a catastrophe before its current age.  The flatness
of space is a consequence of the same scaling law, because for large time scales the curvature may be neglected
as it is multiplied by a factor of $k^2$.  Finally, observe that the stationary point at $(1,0)$ corresponds to
singularity in $\widetilde g|_{s(M)}$. Notice that if $x^2=1$, then $S'(t)=0$.  At such points $Ts(M)$ is tangent
to the null cone of $\widetilde g$ and so space-time at the singularity does not have a well defined spatial
metric.

\section{Submanifold geometry for a geodesic observer field}

In the previous section the implications of EGR are studied for the simplest interesting example, namely Friedmann
geometries. In general it is difficult to compute the intrinsic geometry of an arbitrary submanifold $s(M)$ of $T^*M$.
Although general formulas for the curvature of such submanifolds can be developed, they are not at this point
particularly illuminating and so will not be fully exposed here. Rather, this section will demonstrate that if $s$ has
constant length and $\ell^{-1}(s)$ is a geodesic vector field, then the difference between the curvature on $s(N)$
induced by $\pi^*g$ and $\widetilde g$ depends only on the first three covariant derivatives of $s$, and can be
developed in a power series in $\epsilon$ with first non-vanishing term of order $\epsilon^2$. Thus, the additional
terms that appear in the Einstein tensor when $\widetilde g$ is used in the place of $\pi^*g$ can be viewed as
additions to energy momentum tensor that are coupled to gravity by higher powers of the gravitational constant.

To understand the intrinsic curvature of $s(M)$, first calculate the Levi-Civita connection $\widetilde \nabla$ of
$\widetilde g$  on $T^*M$. Let $R$ be the curvature tensor of the dual Levi-Civita connection in $T^*M$. By
metric duality $R$ determines a tensor $\bar R$ defined on $T^*M\times T^*M\times TM$ with values in $TM$ by the
expression $g(\bar R(s,t)v,w)=g(R(v,w)s,t)$. For vector fields $X$ and $Y$ and 1-form fields $s$ and $t$,
$\widetilde\nabla$ is determined at a point $p \in T^*M$ by the expressions
\begin{eqnarray*}
{\widetilde\nabla_{\widetilde X}\widetilde
Y}_p&=&\widetilde{\nabla_XY}+{1\over\epsilon}g(p,p)g(X,Y)i^{-1}p-i^{-1}{1\over 2}R(X,Y)p \\
{\widetilde\nabla_{i^{-1}t}\widetilde Y}_p&=&{g(y,p)\over g(p,p)}\widetilde X - {\epsilon^2\over
2g(p,p)^2}\widetilde{\bar R(p,t)X} \\
\widetilde\nabla_{\widetilde X}i^{-1}t_p&=&i^{-1}\nabla_Xt+{g(t,p)\over g(p,p)}\widetilde X- {\epsilon^2\over
2g(p,p)^2}\widetilde{\bar R(p,t)X} \\
{\widetilde\nabla_{i^{-1}t}i^{-1}s}_p&=&-{1\over g(p,p)}(g(p,t)i^{-1}s+g(p,s)i^{-1}t-g(s,t)i^{-1}p).
\end{eqnarray*}

Now fix a section $s$ of $T^*M$ that has constant length and is closed. Since $\nabla_{\cdot}s$ is a bundle
morphism between $TM$ and $T^*M$, define $\nabla_\cdot s^T$ to be its transpose; that is, $\nabla_\cdot s^T$ is a
morphism between $T^*M$ and $TM$. The conditions that $s$ be of constant length and closed can be stated in the
form $\nabla_ss^T=0$ and $\nabla_Us=\ell\nabla_{\ell U}s^T$ where $U$ is a vector field on $M$ and $\ell\colon TM
\to T^*M$ is the identification map determined by the metric $g$. Note that these identities immediately imply
the classical result that $\ell^{-1}s$ is a geodesic vector field.  If $s$ is assumed to be time-like and
$m^2=-g(s,s)$ then for vector fields $U$ and $W$ on $M$ the above expressions for $\widetilde\nabla$ imply that
\begin{eqnarray*}
\widetilde\nabla_{s_*U}s_*W&=&\widetilde\nabla_{\widetilde U+i^{-1}\nabla_Us}\widetilde W+i^{-1}\nabla_Ws \\
                            &=&\widetilde{\nabla_UW}+{\epsilon^2\over2m^4}\left(\widetilde{\bar R(s,\nabla_U)W}+\widetilde{\bar R(s,\nabla_W)U}\right)-{1\over 2}i^{-1}R(U,W)s+ \\
                            & &\qquad i^{-1}\nabla_U\nabla_Ws - \left({m^2\over\epsilon^2}g(U,W)+{1\over
                            m^2}g(\nabla_Us,\nabla_Ws)\right)i^{-1}s.
\end{eqnarray*}
After some calculation using this form for $\widetilde\nabla_{s_*U}s_*W$, one obtains the following expression for
the sectional curvature $\widetilde K(s_*U,s_*W)=\widetilde g(\widetilde R(s_*U,s_*W)s_*U,s_*W)$ of $\widetilde g$
along $s(N)$
\begin{eqnarray*}
\lefteqn{\widetilde K(s_*U,s_*W)=} \\
                        & &-{m^2\over \epsilon}K(U,W)+{\epsilon\over m^2}\Big(g(\nabla_UR(U,W)s,\nabla_Ws)+\\
                        & &g(\nabla_WR(W,U)s,\nabla_Us) + g(R(U,W)\nabla_Us,\nabla_Ws)\Big) + \\
                        & &{3\epsilon\over 4m^2}g(R(U,W)s,R(U,W)s)+{\epsilon^3\over 4m^4}\Big(g(\bar
                        R(s,\nabla_Us)W,\bar R(s,\nabla_Ws)U) + \\
                        & & g(\bar R(s,\nabla_Us)W,\bar R(s,\nabla_Us)W) + g(\bar R(s,\nabla_Ws)U,\bar R(s,\nabla_Ws)U)- \\
                        & & 3g(R(s,\nabla_Us)U,\bar R(s,\nabla_Ws)W)\Big)+ {\epsilon\over m^4}\Big(g(\nabla_Us,\nabla_Ws) ^2 -\\
                        & &g(\nabla_Us,\nabla_Us) g(\nabla_Ws,\nabla_Ws)\Big)-{m^4\over
                        \epsilon^3}\Big(g(U,W)^2-g(U,U)g(W,W)\Big).
\end{eqnarray*}

To obtain the intrinsic sectional curvature of $s(M)$ from this expression requires the second fundamental form of
$s(M)$, $\alpha(U,W)= {P'}^\bot(\widetilde\nabla_{s_*U}s_*W)$. The orthogonal projection ${P'}^\bot$ on to the compliment
of $Ts(M)$ can be expressed in terms of covariant derivatives of $s$. To see this note that the tangent space to $s(N)$
and its compliment are described as the graphs of the endomorphisms $C\colon H_s\to VT^*M_s$ and $C^*\colon VT^*M_s\to
H_s$; that is, $Ts(M)_s=H_s\oplus C(H_s)$ and $Ts(M)^\bot_s=VT^*M_s\oplus C(VT^*M_s)$. It is easy to see that for a
vector field $U$, $C(\widetilde U)=i^{-1}\nabla_Us$ and $C^*(i^{-1}t)=(\epsilon^2/m^4)\widetilde{\nabla_ts^T}$. The
projection  ${P'}^\bot$  can be expressed in terms of  $C$ and $C^*$, and the orthogonal projections $P$ and $P^\bot$
that map $TT^*M$ onto $H$ and its orthogonal compliment $VT^*M$, and has the form
\[
{P'}^\bot=(P^\bot+C^*P^\bot)(P^\bot-CC^*P^\bot)^\#(P^\bot-CP)
\]
where $\strut^\#$ denotes the group inverse.

The assumption that $s$ is of constant length and closed implies that $\ell\nabla_\cdot s$ has the structure of a
3-dimensional symmetric linear map $A$, and hence satisfies the polynomial identity $A^3=\xi_1A^2-\xi_2A+\xi_3$.
The fundamental invariants $\xi_1$, $\xi_2$ and $\xi_3$ are given in terms of the traces of powers of $A$ by the
expressions
\begin{eqnarray*}
\xi_1&=&\mathrm{tr}(A) \\
\xi_2&=&{1\over 2}(\mathrm{tr}(A)^2-\mathrm{tr}(A^2)) \\
\xi_3&=&{1\over 6}\mathrm{tr}(A)^3-{1\over 2}\mathrm{tr}(A)\mathrm{tr}(A^2)+{1\over 3}\mathrm{tr}(A)^3.
\end{eqnarray*}
These invariants may be used to reduce $(1-A^2)^{-1}$ to the expression
\begin{eqnarray*}
\lefteqn{(1-A^2)^{-1}=}   \\
         & &{1+\xi_1-2\xi_2+\xi^2\over 1+\xi_1^2-2\xi_2+\xi_1\xi_3+{1\over 3}\xi_1^3\xi_3-{2\over
3}\xi_1\xi_2\xi_3+{1\over 3}\xi_3^2}+  \\
        & & {\xi_3-\xi_1\xi_2\over 1+\xi_1^2-2\xi_2+\xi_1\xi_3+{1\over
3}\xi_1^3\xi_3-{2\over 3}\xi_1\xi_2\xi_3+{1\over 3}\xi_3^2}A+    \\
        & & {1+2\xi_1^2+3\xi_2\over 1+\xi_1^2-2\xi_2+\xi_1\xi_3+{1\over 3}\xi_1^3\xi_3-{2\over 3}\xi_1\xi_2\xi_3+{1\over
3}\xi_3^2}A^2.
\end{eqnarray*}

To use these identities to write an expression for ${P'}^\bot$ in closed form, introduce $\widetilde\ell\colon H\to
VT^*M$ defined by $\widetilde\ell(\widetilde U)=i^{-1}\ell(U)$, and endomorphisms $B^*$ and $B$ of $VT^*M$ and
$H$ given by $B^*(i^{-1}t)=i^{-1}\nabla_{\ell^{-1}t}s$ and $B(\widetilde U)=\widetilde{\ell^{-1}\nabla_Us}$.  If
$A$ is taken to be either $(\epsilon/m^2)B^*$ or $(\epsilon/m^2)B$ then, since both maps have the same invariants,
${P'}^\bot$ can be seen to be of the form
\begin{eqnarray*}
\lefteqn{{P'}^\bot=}  \\
        & &(1+p_1+p_2{\epsilon\over m^2}B^*+p_3({\epsilon\over m^2}B^*)^2)P^\bot - {m^2\over\epsilon}(q_1+q_2{\epsilon\over m^2}B^*+q_3({\epsilon\over m^2}B^*)^2)\widetilde\ell P \\
        & & + {\epsilon\over m^2}(q_1+q_2{\epsilon\over m^2}B+q_3({\epsilon\over m^2}B)^2)\widetilde\ell^{-1} P^\bot - (p_1+p_2{\epsilon\over m^2}B^*+p_3({\epsilon\over m^2}B^*)^2)P
\end{eqnarray*}
where $p_1$, $p_2$, $p_3$, $q_1$, $q_2$, $q_3$ have explicit rational expressions in terms of the invariants and
have the following orders in $\epsilon$; $p_1=O(\epsilon^4)$, $p_2=O(\epsilon^3)$, $p_3=1+O(\epsilon^2)$,
$q_1=O(\epsilon^3)$, $q_2=1+O(\epsilon^2)$, $q_3=O(\epsilon)$.

The Gauss equation states that if $K_s$ is the intrinsic sectional curvature of $s(N)$, then $K_s$ is related
to the ambient
sectional curvature by
\begin{eqnarray*}
\lefteqn{K_s(s_*U,s_*W)=}  \\
        & &\widetilde K(s_*U,s_*W)-\widetilde g(\alpha(s_*U,s_*W),\alpha(s_*U,s_*W)) +
\widetilde g(\alpha(s_*U,s_*U),\alpha(s_*W,s_*W)).
\end{eqnarray*}
To isolate the terms of order less than $\epsilon$ that are present in this expression first decompose
$\nabla_{s_*U}{s_*W}$ as $\nabla_{s_*U}{s_*W}=T(s_*U,s_*W)+S(s_*U,s_*W)$ where $T$ is vertical and $S$ is
horizontal. Since $\nabla_{s}s^T=0$, it follows that $B^*i^{-1}s=0$, and so the only terms of order less than
$\epsilon$ that contribute to $K_s$ are of the following type
\begin{eqnarray*}
\lefteqn{\widetilde g(\alpha(s_*U,s_*W),\alpha(s_*V,s_*Z))=} \\
        & & \widetilde g(T(s_*U,s_*W),T(s_*V,s_*Z)+ \widetilde
g(T(s_*U,s_*W),B^*\widetilde\ell S(s_*V,s_*Z)) + \\
        & &\widetilde g(B^*\widetilde\ell S(s_*U,s_*W),T(s_*V,s_*Z) +O(\epsilon)\\
        &=& -{m^4\over\epsilon^3}g(U,W)g(V,Z)+O(\epsilon).
\end{eqnarray*}
Hence, the Gauss equation gives that $K_s(s_*U,s_*W)= -(m^2/\epsilon)K(U,W)+O(\epsilon)$. In fact to order
$\epsilon$, a calculation shows that
\begin{eqnarray*}
\lefteqn{K_s(s_*U,S_*W) =} \\
        & &-{m^2\over\epsilon}K(U,W)+{\epsilon\over m^2}(g(R(U,W)s,R(U,W)s) - g(\nabla_WR(U,W)s,\nabla_Us) +\\
        & &g(\nabla_UR(U,W)s,\nabla_Ws) + 3g(R(U,W)\nabla_Us,\nabla_Ws) + \\
        & &g(\nabla_W(\nabla_Us),\nabla_U(\nabla_Ws)) - g(\nabla_W(\nabla_Ws),\nabla_U(\nabla_Us))) +o(\epsilon).
\end{eqnarray*}
Thus the terms that must be added to $K(U,W)$ to obtain $K_s(s_*U,s_*W)$ occur at two order of $\epsilon$ higher
than $K(U,W)$ and involve $\nabla_\cdot s$, $\nabla_\cdot(\nabla_\cdot s)$ and $\nabla_\cdot (\nabla_\cdot
(\nabla_\cdot s))$ as $R(U,W)s=\nabla_U(\nabla_W s)-\nabla_W(\nabla_U s)$.

It is interesting to note that if $GK_s$ is truncated at order $\epsilon^2$, and is used to construct approximate
solutions for Friedmann geometries as in Section 2, then the solution curve $J_0$ does not return to the
infationary state but rather recedes to infinity in backward finite time for all values of $\alpha$.

\end{document}